\documentclass[pdftex,twocolumn,epjc3]{svjour3}
\RequirePackage{graphicx}
\RequirePackage{amsmath}
\RequirePackage{amssymb}
\RequirePackage[numbers,sort&compress]{natbib}
\RequirePackage[colorlinks,citecolor=blue,urlcolor=blue,linkcolor=blue]{hyperref}
\journalname{European Physical Journal B}
\begin{document}

\title{Tempered fractional Brownian motion on finite intervals %\thanksref{t1}
}


\author{Thomas Vojta \thanksref{e1,addr1}    \and
        Zachary Miller  \thanksref{e2,addr1}   \and
        Samuel Halladay \thanksref{e3,addr1,addr2}
}

\thankstext{e1}{e-mail: vojtat@mst.edu}
\thankstext{e2}{e-mail: zamydm@mst.edu}
\thankstext{e3}{e-mail: samuel.halladay@yale.edu}

\institute{Department of Physics, Missouri University of Science and Technology, Rolla, Missouri 65409, USA \label{addr1}
\and
Department of Applied Physics, Yale University, New Haven, Connecticut 06520, USA \label{addr2}
}

\date{Received: date / Accepted: date}

\maketitle

\begin{abstract}
Diffusive transport in many complex systems features a crossover between anomalous diffusion at short times and normal
diffusion at long times. This behavior can be mathematically modeled by cutting off (tempering) beyond a mesoscopic
correlation time the power-law correlations between the increments of fractional Brownian motion. Here, we investigate
such tempered fractional Brownian motion confined to a finite
interval by reflecting walls. Specifically, we explore how the tempering of the long-time correlations affects the
strong accumulation and depletion of particles near reflecting boundaries recently discovered for untempered fractional
Brownian motion. We find that exponential tempering introduces a characteristic size for the accumulation and depletion
zones but does not affect the functional form of the probability density close to the wall. In contrast, power-law
tempering leads to more complex behavior that differs between the superdiffusive and subdiffusive cases.
% \keywords{First keyword \and Second keyword \and More}
\end{abstract}

%%%%%%%%%%%%%%%%%%%%%%%%%%%%%%%%%%%%%%%%%%%%%%%%%%%%%%%%%%%%%%%%%%%%%%%%%%%%%%%%%%%%%%%%%%%%%%%%%%%%%%
\section{Introduction}
\label{sec:intro}
%%%%%%%%%%%%%%%%%%%%%%%%%%%%%%%%%%%%%%%%%%%%%%%%%%%%%%%%%%%%%%%%%%%%%%%%%%%%%%%%%%%%%%%%%%%%%%%%%%%%%%

Diffusive transport phenomena can be found in a wide variety of fields such as physics, chemistry, biology, and beyond.
According to Einstein \cite{Einstein_book56}, Langevin \cite{Langevin08}, and Smoluchowski \cite{Smoluchowski18},
diffusion arises from the motion of the particles in question being stochastic. It is often characterized in terms of
the power-law relation $\langle x^2 \rangle \sim t^\alpha$ between the mean-square displacement $\langle x^2 \rangle$
of a diffusing particle and the elapsed time $t$. The exponent value $\alpha=1$ corresponds to normal diffusion
which emerges naturally if the stochastic motion is local in time and space \cite{Hughes95}.
Recently, there has been significant interest in stochastic motion with $\alpha \ne 1$, i.e., in anomalous
diffusion \cite{BouchaudGeorges90,MetzlerKlafter00}. Depending on the value of the anomalous diffusion exponent
$\alpha$, one can distinguish subdiffusion ($0 < \alpha < 1$) for which $\langle x^2 \rangle$ grows
slower than $t$ and superdiffusion ($1 < \alpha < 2$) for which $\langle x^2 \rangle$ grows faster than $t$. Both subdiffusion
and superdiffusion have been observed experimentally in numerous systems (see. e.g.\ Ref.\
\cite{HoeflingFranosch13,BressloffNewby13,MJCB14,MerozSokolov15,MetzlerJeonCherstvy16,Norregaardetal17}
and references therein), in part  because modern microscopy provides unprecedented information about the motion of
single molecules in complex environments \cite{XCLLL08,BrauchleLambMichaelis12,ManzoGarciaParajo15}.

Anomalous diffusion can arise if the random motion violates the condition of locality in time and space, e.g.,
when individual displacements (steps) of the diffusing particle are long-range correlated in time.
Fractional Brownian motion (FBM) is a paradigmatic mathematical model of this situation. It was was first
introduced by Kolmogorov \cite{Kolmogorov40} and later explored by Mandelbrot and van Ness \cite{MandelbrotVanNess68}.
FBM is a self-similar Gaussian stochastic process with long-time (power-law) correlated increments
which are antipersistent (anticorrelated) in the subdiffusive regime, $0<\alpha<1$, but persistent (positively
correlated) in the superdiffusive regime $1<\alpha<2$. In the marginal case, $\alpha=1$, FBM is identical to normal
Brownian motion with uncorrelated increments.
FBM processes have been used to describe the motion inside biological cells
\cite{SzymanskiWeiss09,MWBK09,WeberSpakowitzTheriot10,Jeonetal11,JMJM12,Tabeietal13}, the patterns of serotonergic fibers
in vertebrate brains \cite{JanusonisDetering19,JanusonisDeteringMetzlerVojta20},
polymer dynamics \cite{ChakravartiSebastian97,Panja10}, electronic network traffic \cite{MRRS02}, as well as
fluctuations of financial markets \cite{ComteRenault98,RostekSchoebel13}.

Even though FBM has been explored quite extensively in mathematical literature (see, e.g., Refs.\
\cite{Kahane85,Yaglom87,Beran94,BHOZ08}), much of its behavior in confined geometries remains elusive
because a generalized diffusion equation for FBM has yet to be found. Additionally, the method of images
\cite{MetzlerKlafter00,Redner_book01}, often invoked for boundary value problems, fails.
Existing results concern the first-passage problem on a semi-infinite interval
\cite{HansenEngoyMaloy94,DingYang95,KKMCBS97,Molchan99}) and two-dimensional wedge
and parabolic domains \cite{JeonChechkinMetzler11,AurzadaLifshits19}. In addition, properties of FBM close to an
absorbing boundary were investigated in Refs.\
\cite{ChatelainKantorKardar08,ZoiaRossoMajumdar09,WieseMajumdarRosso11,DelormeWiese15,DelormeWiese16,ArutkinWalterWiese20,VojtaWarhover21}.

Recently, reflected FBM has attracted significant attention because the interplay between the long-time
correlations and the reflecting barriers modifies the probability density function $P(x,t)$ of the diffusing particles.
In the case of superdiffusive FBM, particles accumulate at the barrier whereas they are depleted near
the barrier for subdiffusive FBM. More specifically, on a semi-infinite interval with a reflecting wall at
the origin, $P$ becomes highly non-Gaussian and develops a power-law singularity, $P \sim x^\kappa$, at the wall \cite{WadaVojta18,WadaWarhoverVojta19}. On a finite interval with reflecting walls at both ends, the stationary probability density
deviates from the uniform distribution found for normal diffusion \cite{Guggenbergeretal19} and also
features power-law singularities at the walls \cite{VHSJGM20}. Analogous results were obtained in higher dimensions
\cite{VHSJGM20}.

In many of the experimental systems that feature anomalous diffusion, the anomalous power law  $\langle x^2 \rangle \sim t^\alpha$
with $\alpha \ne 1$ does not extend to arbitrarily long times but eventually crosses over to normal diffusion ($\alpha=1$)
when the time exceeds a characteristic correlation time. To model this crossover, Molina-Garcia et al.
\cite{MolinaGarciaetal18} introduced the notion of
tempered FBM, a stochastic process in which the long-time power-law correlations are cutoff beyond the tempering time
$t_*$.\footnote{A different type of tempering was proposed by Meerschaert and Sabzikar \cite{MeerschaertSabzikar13}. It leads
to fundamentally different behavior and does not describe the anomalous to normal diffusion crossover. We will briefly come back to
this point in the concluding section.}
As the unusual behavior of the probability density of reflected FBM stems from the interplay of the reflecting barriers and the
long-time correlations, it is important to ask how the tempering of these correlations affects the probability density.

Here, we therefore study the behavior of tempered FBM that is confined to a finite interval by reflecting walls
at both ends. We employ large-scale computer simulations to study the mean-square displacement as well as the probability density
function for hard exponential tempering of the correlations as well as softer power-law tempering. We distinguish the superdiffusive
and subdiffusive regimes and compare our findings to the corresponding behavior of untempered FBM.

Our paper is organized as follows. We introduce FBM and tempered FBM in Sec.\ \ref{sec:FBM}. Section
\ref{sec:rFBM_review} briefly summarizes key properties of (untempered) FBM with reflecting walls for later comparison
with the tempered case. Simulation results for exponentially tempered FBM on a finite interval with reflecting walls
at both ends are presented in Sec.\ \ref{sec:results_exp} whereas the corresponding results for power-law
tempering are shown in Sec.\ \ref{sec:results_pow}. We conclude in Sec.\ \ref{sec:conclusions}.

%%%%%%%%%%%%%%%%%%%%%%%%%%%%%%%%%%%%%%%%%%%%%%%%%%%%%%%%%%%%%%%%%%%%%%%%%%%%%%%%%%%%%%%%%%%%%%%%%%%%%%
\section{Fractional Brownian motion and tempered fractional Brownian motion}
\label{sec:FBM}
%%%%%%%%%%%%%%%%%%%%%%%%%%%%%%%%%%%%%%%%%%%%%%%%%%%%%%%%%%%%%%%%%%%%%%%%%%%%%%%%%%%%%%%%%%%%%%%%%%%%%%
\subsection{Definition of fractional Brownian motion}
\label{subsec:FBM_definition}
%%%%%%%%%%%%%%%%%%%%%%%%%%%%%%%%%%%%%%%%%%%%%%%%%%%%%%%%%%%%%%%%%%%%%%%%%%%%%%%%%%%%%%%%%%%%%%%%%%%%%%

We start from the definition of FBM as a continuous-time centered Gaussian stochastic
process. Consider a particle located at position $X=0$ at time $t=0$. The covariance function of its position
$X$ at later times $s$ and $t$ is given by
\begin{equation}
\langle X(s) X(t) \rangle = K (s^\alpha - |s-t|^\alpha + t^\alpha)
\label{eq:FBM_cov}
\end{equation}
where the exponent $\alpha$ is in the range $0 < \alpha < 2$.
Setting $s=t$ results in a mean-square displacement of $\langle X^2 \rangle = 2 K t^\alpha$, i.e.,
the particle undergoes anomalous diffusion, with $\alpha$ playing the role of the anomalous diffusion
exponent.

In preparation of the computer simulations, we now discretize time, $t_n= \epsilon n$,
and define positions  $x_n = X(t_n)$. Here, $\epsilon$ is the time step, and $n$ is an integer. The resulting
discrete version of FBM \cite{Qian03} can be understood as a random walk with identically
Gaussian distributed and long-time correlated steps. The particle position $x_n$ now evol\-ves according
to the recursion relation
\begin{equation}
x_{n+1} = x_n + \xi_n~.
\label{eq:FBM_recursion}
\end{equation}
Here, the increments $\xi_n$ constitute a discrete fractional Gaussian noise, a stationary
Gaussian process of zero mean, variance $\sigma^2 = 2 K \epsilon^\alpha$, and covariance
\begin{equation}
C_n^\textrm{FBM}=\langle \xi_m \xi_{m+n} \rangle = \frac 1 2 \sigma^2 (|n+1|^\alpha - 2|n|^\alpha + |n-1|^\alpha)~.
\label{eq:FGN_cov}
\end{equation}
The covariance is positive (persistent) for $\alpha>1$ and negative (anti-persistent)
for $\alpha < 1$ (and $n\ne 0$). If $\alpha=1$, the covariance vanishes for all $n\ne 0$ leading to an
uncorrelated random walk,
i.e., normal Brownian motion. In the long-time limit $n\to \infty$, the covariance follows
the power-law form $\langle \xi_m \xi_{m+n} \rangle  \sim\alpha (\alpha-1) |n|^{\alpha-2}$.

The time discretization error becomes unimportant if the time step $\epsilon$ is small compared to the considered
times $t$. Equivalently, the individual step size $\sigma$ needs to be small compared to the considered distances
or system sizes.
This continuum limit can be reached either by taking the time step $\epsilon$ to zero at fixed total time $t$ or,
equivalently, by taking $t$ to infinity at fixed $\epsilon$. We will follow the latter route by setting
$\epsilon=\mathrm{const}$ and considering long times $t \to \infty$.

%%%%%%%%%%%%%%%%%%%%%%%%%%%%%%%%%%%%%%%%%%%%%%%%%%%%%%%%%%%%%%%%%%%%%%%%%%%%%%%%%%%%%%%%%%%%%%%%%%%%%%
\subsection{Tempering the correlations}
\label{subsec:tempering}
%%%%%%%%%%%%%%%%%%%%%%%%%%%%%%%%%%%%%%%%%%%%%%%%%%%%%%%%%%%%%%%%%%%%%%%%%%%%%%%%%%%%%%%%%%%%%%%%%%%%%%

To model the crossover between anomalous diffusion and normal diffusion that is observed in many experimental systems,
we now follow Ref.\ \cite{MolinaGarciaetal18} and introduce a tempering (truncation) of the long-range correlations
encoded in the covariance (\ref{eq:FGN_cov}) of the fractional Gaussian noise.
We will consider both a ``hard'' exponential tempering and a ``softer'' power-law tempering.

In the case of exponential tempering, the noise covariance (\ref{eq:FGN_cov}) gets replaced by
\begin{equation}
C_n = C_n^\textrm{FBM} \exp(-|t_n|/t_*)
\label{eq:FGN_cov_exp}
\end{equation}
where $t_*$ is the tempering (crossover) time scale governing the crossover from anomalous diffusion
for times less than $t_*$ to normal diffusion on time scales larger than $t_*$. For power-law tempering,
the noise covariance reads
\begin{equation}
C_n = C_n^\textrm{FBM}  ( 1+ |t_n| / t_*)^{-\mu}~.
\label{eq:FGN_cov_pow}
\end{equation}
It is characterized by a positive decay exponent $\mu$ in addition to the tempering time $t_*$.
Note that the Fourier transform $\tilde C(\omega)$ of the covariance $C_n$ must be nonnegative because
it represents the power spectrum of the noise $\xi_n$. Both (\ref{eq:FGN_cov_exp}) and (\ref{eq:FGN_cov_pow})
fulfill this condition as was demonstrated in Ref.\ \cite{MolinaGarciaetal18} and verified numerically
in our simulations.

\begin{figure}
\includegraphics[width=\columnwidth]{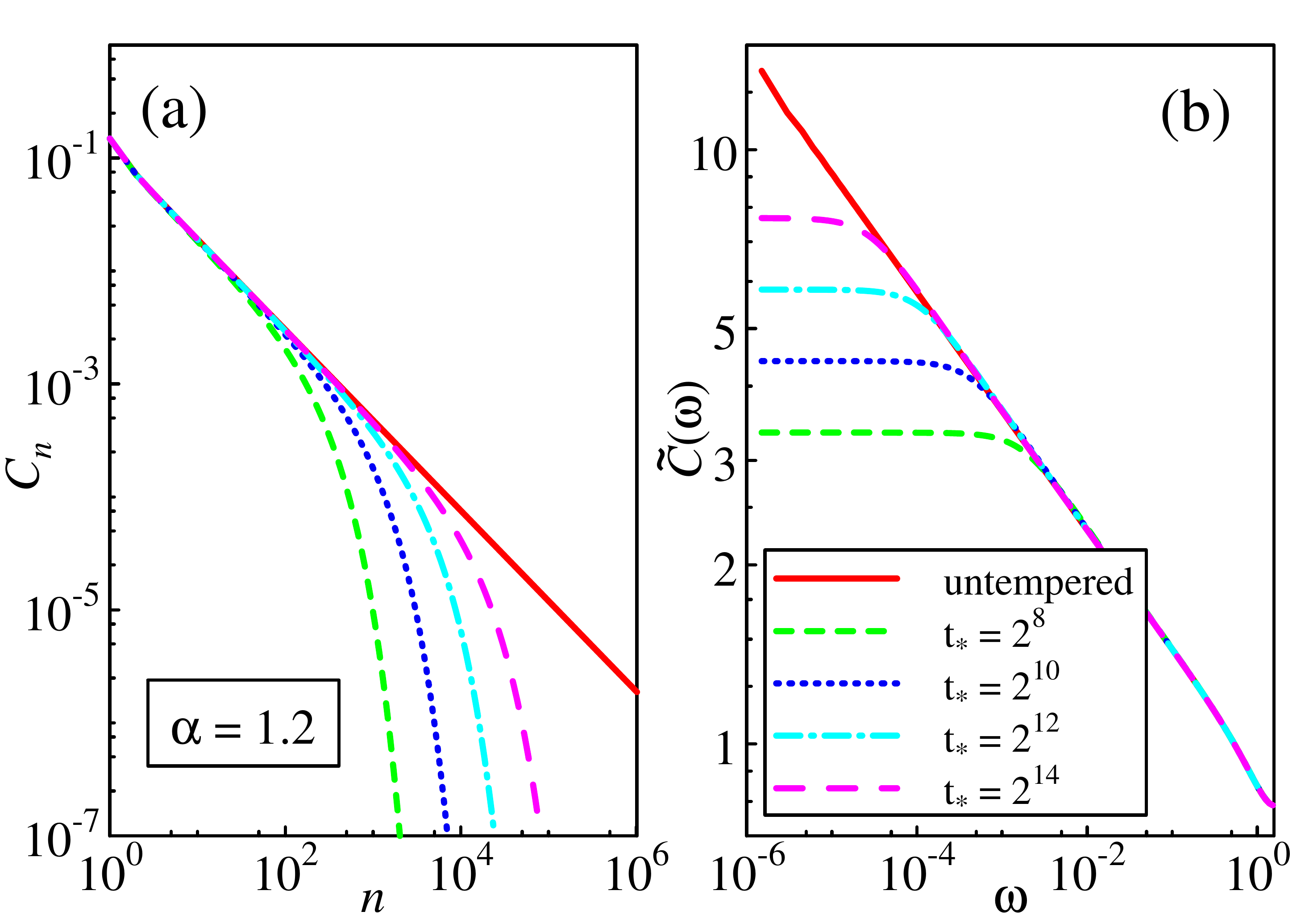}
\caption{(a) Covariance $C_n$ of exponentially tempered fractional Gaussian noise with $\alpha=1.2$ for different
values of the tempering time $t_*$. (b) Corresponding Fourier transforms $\tilde C(\omega)$, representing the power
spectra of the noise.}
\label{fig:covariance_gamma08}
\end{figure}

\begin{figure}
\includegraphics[width=\columnwidth]{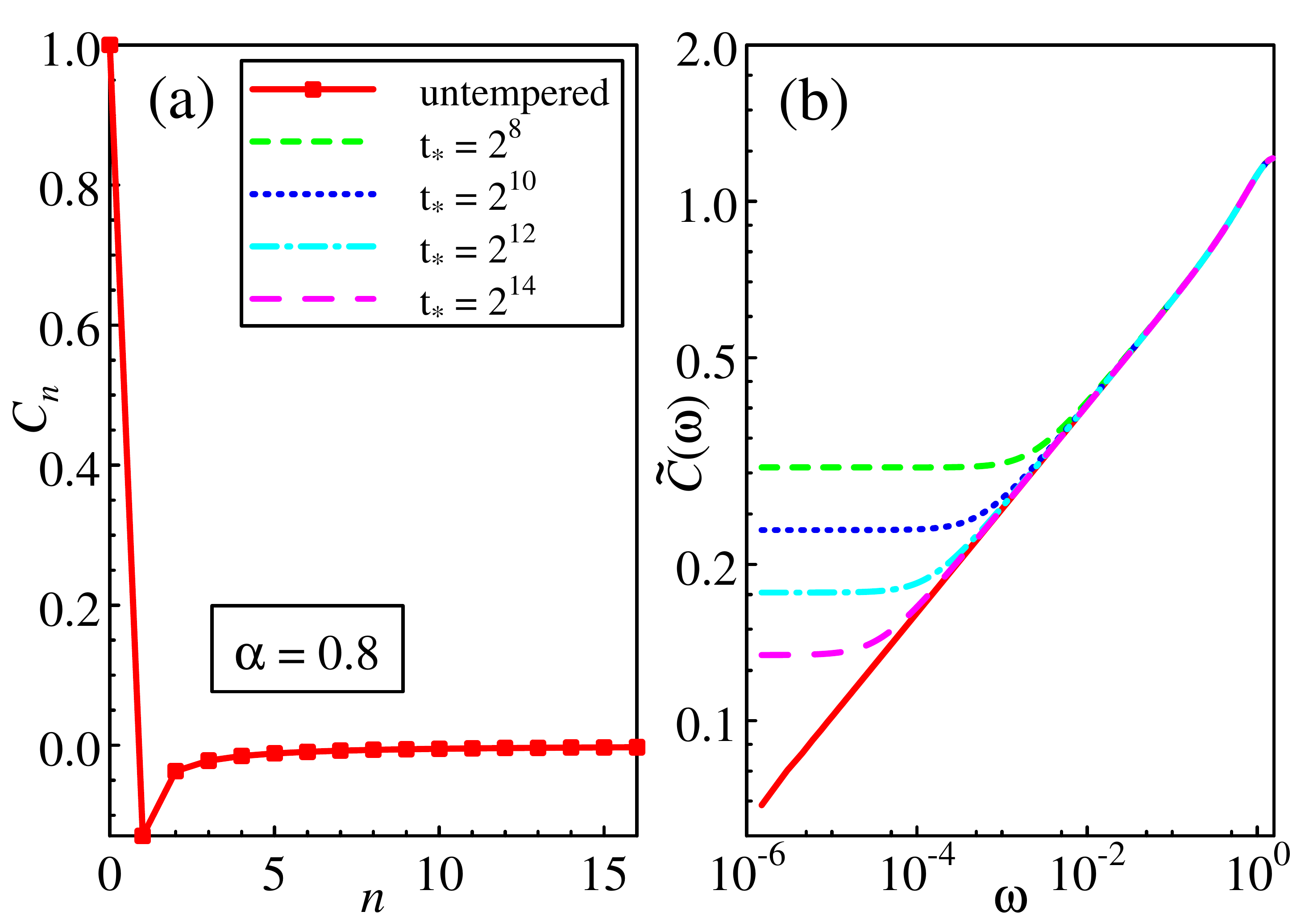}
\caption{(a) Covariance $C_n$ of exponentially tempered fractional Gaussian noise with $\alpha=0.8$ for different
values of the tempering time $t_*$. The time axis in the plot is restricted to $n \le 15$ to make the negative (anti) correlations
clearly visible. For these $n$, the curves for all studied tempering times $t_*$ coincide.
(b) Corresponding Fourier transforms $\tilde C(\omega)$, representing the power
spectra of the noise.}
\label{fig:covariance_gamma12}
\end{figure}

Exponentially tempered fractional Gaussian noise with $\alpha=1.2$ (in the superdiffusive
regime) is illustrated in Fig.\ \ref{fig:covariance_gamma08}.
Panel (a) shows how the power-law correlations are truncated beyond the tempering time. The corresponding noise power spectra
in panel (b) are indeed nonnegative and feature crossovers from the FBM power law $\tilde C \sim \omega^{1-\alpha}$ at
higher frequencies to $\tilde C = \textrm{const}$
at lower frequencies when the noise becomes effectively uncorrelated.
Figure \ref{fig:covariance_gamma12}(a) illustrates the negative (anti-persistent) covariance for $\alpha=0.8$ (in the
subdiffusive regime).
The corresponding Fourier transforms $\tilde C(\omega)$, shown in Fig.\ \ref{fig:covariance_gamma12}(b) for several tempering times,
cross over from $\tilde C \sim \omega^{1-\alpha}$ to $\tilde C = \textrm{const}$ just as in the superdiffusive case.

\begin{figure}
\includegraphics[width=\columnwidth]{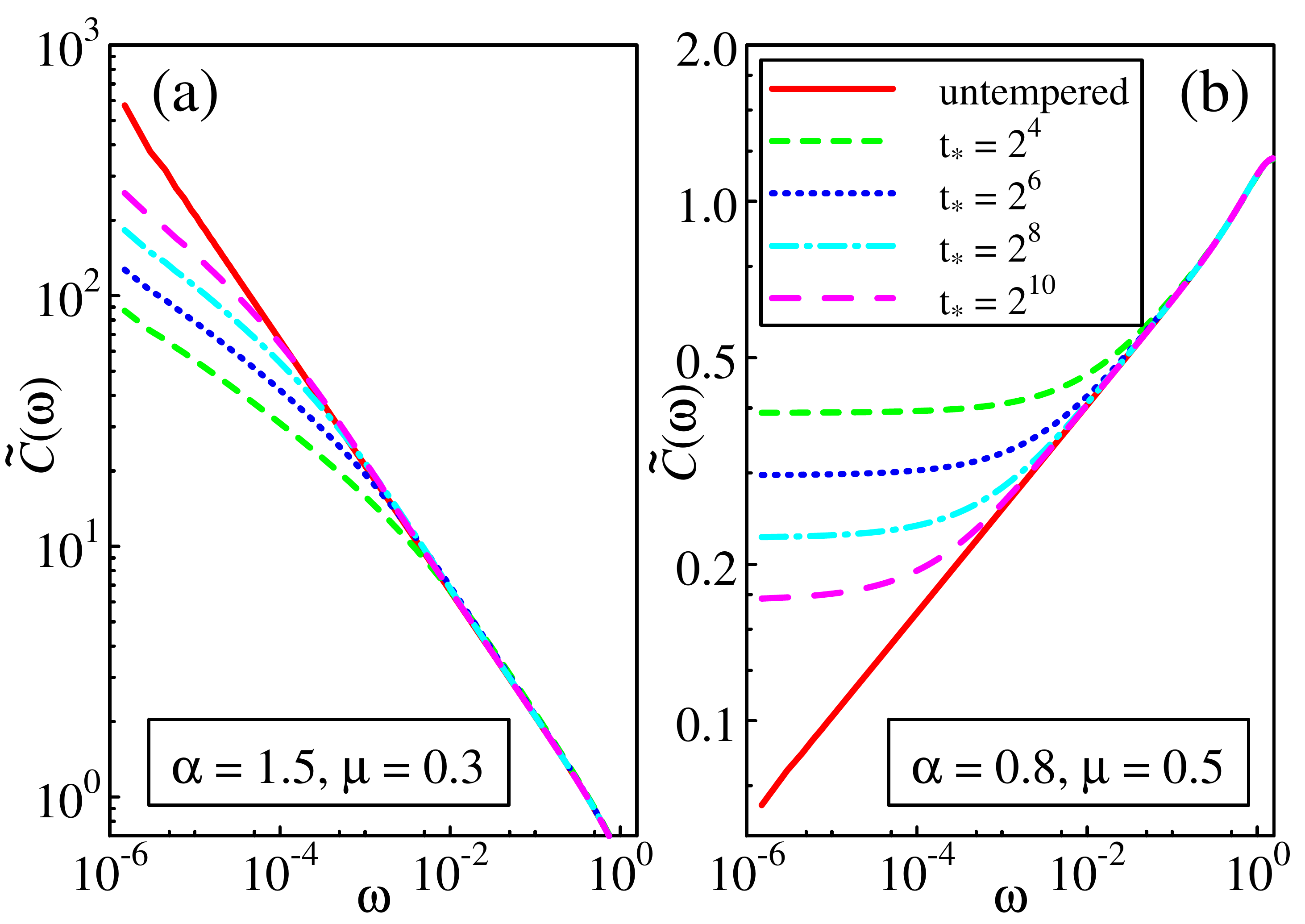}
\caption{Power spectrum $\tilde C(\omega)$ of power-law tempered fractional Gaussian noise for different values of the tempering time $t_*$.
(a) $\alpha=1.5$ and $\mu=0.3$  (b) $\alpha=0.8$ and $\mu=0.5$.}
\label{fig:covariance_gamma05_12_pow}
\end{figure}

The effects of power-law tempering are more complex than those of exponential tempering, and they differ between the
superdiffusive and subdiffusive regimes. Let us first consider superdiffusive FBM ($1< \alpha < 2$). In the presence
of power-law tempering, the asymptotic large-$n$ behavior of the noise covariance (\ref{eq:FGN_cov_pow}) is given by
$C_n \sim |n|^{\alpha-2-\mu}$.  If $\alpha-2-\mu > -1$ (called ``weak power-law tempering'' in Ref.\ \cite{MolinaGarciaetal18}),
the Fourier transform $\tilde C (\omega)$ of the covariance diverges
as $\tilde C \sim \omega^{1-\alpha+\mu}$
for $\omega \to 0$ implying that the power-law correlations are still relevant [see Fig.\ \ref{fig:covariance_gamma05_12_pow}(a)].
If $\alpha-2-\mu < -1$ (called ``strong power-law tempering'' in Ref.\ \cite{MolinaGarciaetal18}),
the Fourier transform $\tilde C (\omega)$ of the covariance approaches a constant for $\omega \to 0$
as in the case of uncorrelated disorder.

Let us now turn to the subdiffusive case ($0 < \alpha < 1$). The Fourier transform $\tilde C (\omega)$ of untempered
fractional Gaussian noise vanishes for $\omega \to 0$ in this regime, reflecting the perfect anticorrelations, $\sum_n C_n = 0$,
of the noise.
As any tempering destroys this equality (unless $\sum_n C_n$ is fine tuned to zero),
the power spectra of the tempered noise contain an uncorrelated
component reflected in the nonzero low-frequency limit of the Fourier transform $\tilde C(\omega)$,
see Fig.\ \ref{fig:covariance_gamma05_12_pow}(b).

%%%%%%%%%%%%%%%%%%%%%%%%%%%%%%%%%%%%%%%%%%%%%%%%%%%%%%%%%%%%%%%%%%%%%%%%%%%%%%%%%%%%%%%%%%%%%%%%%%%%%%
\subsection{Reflecting boundaries}
\label{subsec:reflecting_boundaries}
%%%%%%%%%%%%%%%%%%%%%%%%%%%%%%%%%%%%%%%%%%%%%%%%%%%%%%%%%%%%%%%%%%%%%%%%%%%%%%%%%%%%%%%%%%%%%%%%%%%%%%

Reflecting boundaries that confine the motion of the diffusing particle can be introduced by modifying
the recursion (\ref{eq:FBM_recursion}). The fractional Gaussian noise $\xi_n$ defining the increments
is understood as externally given \cite{Klimontovich_book95}; it is therefore not modified by the barriers.
Different implementations of the reflecting boundary conditions and their effects on FBM were studied
in Ref.\ \cite{VHSJGM20}. This paper demonstrated that details of the wall implementation
are unimportant in the continuum limit. They influence the behavior only in a narrow spatial region
close to wall (whose size is controlled by the step size $\sigma$).

Here, we define a reflecting boundary at position $w$ that restricts the motion to $x \ge w$ by means of the recursion
\begin{equation}
x_{n+1} = \left \{ \begin{array}{ll} x_n+ \xi_n  & \quad \textrm{if} \quad x_n+ \xi_n \ge w\\
                     x_n & \quad \textrm{otherwise}   \end{array}  \right. ~.
\label{eq:FBM_recursion_inelastic}
\end{equation}
In other words, the particle does not move at all if the step would take it into the forbidden region
$x < w$. A reflecting boundary that restricts the motion to $x \le w$ can be defined analogously.

%%%%%%%%%%%%%%%%%%%%%%%%%%%%%%%%%%%%%%%%%%%%%%%%%%%%%%%%%%%%%%%%%%%%%%%%%%%%%%%%%%%%%%%%%%%%%%%%%%%%%%
\subsection{Simulation details}
%%%%%%%%%%%%%%%%%%%%%%%%%%%%%%%%%%%%%%%%%%%%%%%%%%%%%%%%%%%%%%%%%%%%%%%%%%%%%%%%%%%%%%%%%%%%%%%%%%%%%%

In our computer simulations we investigate both exponentially and power-law tempered FBM on a finite interval
of length $L$ with reflecting walls at both ends. We use anomalous diffusion exponents $\alpha$ ranging from
0.6 (in the subdiffusive regime) to 1.6 (in the superdiffusive regime). The time step is set to $\epsilon=1$
and $K=1/2$ which fixes the variance of the individual increments at unity,  $\sigma^2=1$.

Each simulation employs a large number of particles (between 20,000 and more than $10^6$),
leading to small statistical errors (characteristic errors will be given in some of the figure captions).
Each particle carries out up to
$2^{26} \approx 6.7 \times 10^7$ time steps. These long simulation times allow us to reach the
continuum limit for which the time discretization becomes unimportant, as was explained in Sec.\
\ref{subsec:FBM_definition}. Consequently, we select interval lengths that fulfill the condition
$L/\sigma \gg 1$. Specifically, the interval lengths range from $L=500$ for the most
subdiffusive $\alpha$ to $L=10^5$ for the most superdiffusive $\alpha$ values.

The fractional Gaussian noise, i.e., the increments $\xi_n$, are precalculated before each simulation run
using the Fourier-filtering technique \cite{MHSS96}. This method starts from a sequence of independent
Gaussian random numbers $\chi_i$ of zero mean and unit variance (which are created via the Box-Muller
transformation from random numbers produced by the LFSR113  \cite{Lecuyer99} and
KISS 2005 \cite{Marsaglia05}) random number generators.
The Fourier transform $\tilde \chi_\omega$ of these numbers
is then converted via ${\tilde{\xi}_\omega} = [\tilde C(\omega)]^{1/2} \tilde{\chi}_\omega$,
where $\tilde C(\omega)$ is the Fourier transform of the desired covariance function (\ref{eq:FGN_cov_exp})
or (\ref{eq:FGN_cov_pow}).
The inverse Fourier transformation of the ${\tilde{\xi}_\omega}$ gives the desired noise values.

%%%%%%%%%%%%%%%%%%%%%%%%%%%%%%%%%%%%%%%%%%%%%%%%%%%%%%%%%%%%%%%%%%%%%%%%%%%%%%%%%%%%%%%%%%%%%%%%%%%%%%
\section{Review of fractional Brownian motion with reflecting walls}
\label{sec:rFBM_review}
%%%%%%%%%%%%%%%%%%%%%%%%%%%%%%%%%%%%%%%%%%%%%%%%%%%%%%%%%%%%%%%%%%%%%%%%%%%%%%%%%%%%%%%%%%%%%%%%%%%%%%

The behavior of (untempered) fractional Brownian motion in the presence of reflecting boundaries has recently
attracted considerable attention because large-scale computer simulations have demonstrated that the interplay
between the long-time correlations of FBM and the geometric confinement strongly affects the probability density
of the diffusing particles.

In the case of FBM  on the semi-infinite interval $(0,\infty)$ with a reflecting wall at the origin,
particles accumulate close to the wall for superdiffusive FBM ($\alpha >1$) whereas they are depleted
at the wall for subdiffusive FBM ($\alpha < 1$) \cite{WadaVojta18}. Specifically, the probability density
function  $P(x,t)$ of the particle position $x$ at time $t$ develops a power-law singularity, $P \sim x^\kappa$,
for $x \to 0$. Based on extensive numerical data, Wada et al.\ \cite{WadaVojta18} conjectured the relation
$\kappa = 2/\alpha -2$. Analogous results were found for biased FBM on a semi-infinite interval
\cite{WadaWarhoverVojta19}.

The properties of FBM confined to a finite interval by reflecting walls at both ends
were studied in Ref.\ \cite{Guggenbergeretal19}. The computer simulations showed that the
stationary probability density depends on the value of the anomalous diffusion
exponent $\alpha$ and differs from the flat distribution observed for normal diffusion.
More specifically, the stationary probability density $P(x,L)$ on the interval $(-L/2, L/2)$
fulfills the scaling form
\begin{equation}
P(x,L) = \frac 1 L Y_\alpha(x/L)
\label{eq:P_scaling_L}
\end{equation}
in the continuum limit $L \gg \sigma$. Close to the left interval boundary, the $\alpha$-dependent scaling function
$Y_\alpha(z)$ develops a power-law singularity, $Y_\alpha(z) \sim (z+1/2)^\kappa$ governed by the same exponent
$\kappa = {2/\alpha -2}$ as the probability density on the semi-infinite interval \cite{VHSJGM20}.
The behavior near the right interval boundary is analogous.

We emphasize that the accumulation and depletion of the diffusing particles close to reflecting walls arise from the
nonequilibrium nature of FBM. In contrast, the fractional Langevin equation, which is driven by the same fractional
Gaussian noise as FBM but fulfills the fluctuation-dissipation theorem \cite{Kubo66}, reaches a thermal equilibrium
stationary state. The corresponding probability density is governed by the Boltzmann distribution.
This implies a flat probability density on a finite interval with reflecting walls, independent
of the value of $\alpha$, as was confirmed by
computer simulations of the fractional Langevin equation \cite{VojtaSkinnerMetzler19}.

We also note that there is an interesting similarity between the behavior of the probability density
close to a reflecting wall (at position $w$), $P \sim |x-w|^{2/\alpha -2}$, and the corresponding behavior
close to an absorbing wall, $P \sim |x-w|^{2/\alpha -1}$
\cite{ZoiaRossoMajumdar09,WieseMajumdarRosso11,VojtaWarhover21}.

%%%%%%%%%%%%%%%%%%%%%%%%%%%%%%%%%%%%%%%%%%%%%%%%%%%%%%%%%%%%%%%%%%%%%%%%%%%%%%%%%%%%%%%%%%%%%%%%%%%%%%
\section{Results: exponentially tempered fractional Brownian motion}
\label{sec:results_exp}
%%%%%%%%%%%%%%%%%%%%%%%%%%%%%%%%%%%%%%%%%%%%%%%%%%%%%%%%%%%%%%%%%%%%%%%%%%%%%%%%%%%%%%%%%%%%%%%%%%%%%%

In this section, we report the computer simulation results for exponentially tempered FBM,
employing the noise covariance (\ref{eq:FGN_cov_exp}), on the interval $(-L/2,L/2)$ with reflecting
boundaries at both ends.
The particles start from the center of the interval, $x=0$, at time $t=0$.
The simulations proceed until a steady state is reached, i.e., until the mean-square displacement,
the probability density, and other quantities become time-independent.

\begin{figure}
\includegraphics[width=\columnwidth]{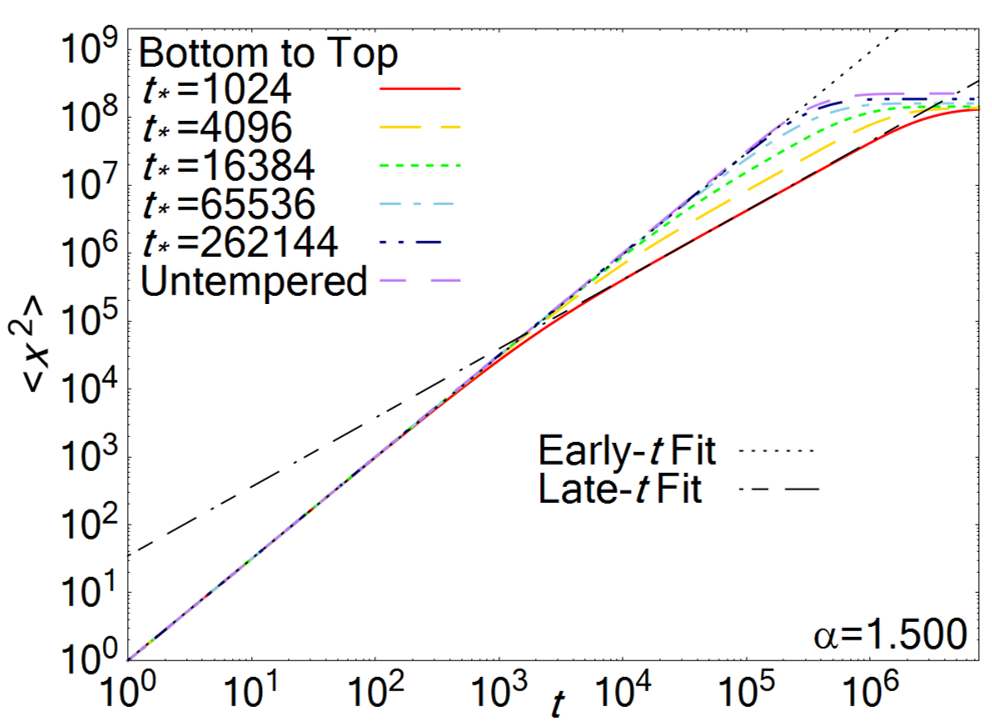}
\caption{Mean-square displacement $\langle x^2 \rangle$ vs.\ time $t$ of exponentially tempered FBM for
interval length $L=40,000$, anomalous diffusion exponent $\alpha=1.5$, and several values of the
tempering time $t_*$. The data are averages over $1.2 \times 10^6$  particles.
The resulting relative statistical error of $\langle x^2 \rangle$ is about $10^{-3}$,
well below the line width. The dotted line is a fit of the
early-time behavior to $\langle x^2 \rangle \sim t^\alpha$
while the dash-dotted line is a fit to normal diffusion $\langle x^2 \rangle \sim t$.}
\label{fig:msd_alpha150}
\end{figure}

\begin{figure}
\includegraphics[width=\columnwidth]{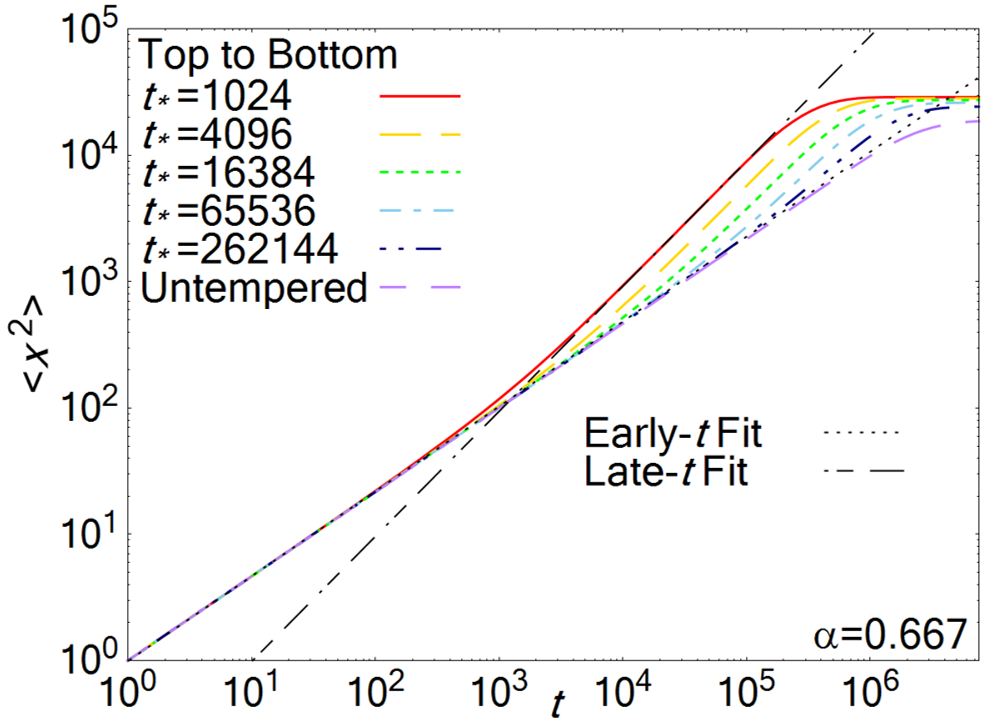}
\caption{Mean-square displacement $\langle x^2 \rangle$ vs.\ time $t$ of exponentially tempered FBM for
interval length $L=600$, anomalous diffusion exponent $\alpha=0.667$, and several values of the
tempering time $t_*$. The data are averages over $1.2 \times 10^6$ particles.
The dotted line is a fit of the early-time behavior to $\langle x^2 \rangle \sim t^\alpha$
while the dash-dotted line is a fit to normal diffusion $\langle x^2 \rangle \sim t$.}
\label{fig:msd_alpha0667}
\end{figure}

To make contact with Ref.\ \cite{MolinaGarciaetal18} where tempered FBM was introduced, we first
discuss the time evolution of the mean-square displacement $\langle x^2 \rangle$. Figure \ref{fig:msd_alpha150}
presents $\langle x^2 \rangle$ as a function of time $t$ for the case of the
superdiffusive anomalous diffusion exponent $\alpha=1.5$ and several tempering times $t_*$.
The data clearly reveal three different time regimes. Initially, for times small compared to the tempering time $t_*$,
the mean-square displacement follows the same anomalous diffusion law $\langle x^2 \rangle \sim t^\alpha$
as untempered (and unconfined) FBM. When the time reaches $t_*$, the mean-square displacement undergoes a
sharp crossover to normal diffusion $\langle x^2 \rangle \sim t$. Finally, $\langle x^2 \rangle$
saturates at a time-independent value indicating that a steady state has been reached. (Note that for
a sufficiently large tempering time, $\langle x^2 \rangle$ may saturate before reaching the crossover to normal
diffusion.)
The properties of subdiffusive tempered FBM are completely analogous, as can be seen in Fig.\
\ref{fig:msd_alpha0667} which presents the time evolution of the mean-square displacement for
$\alpha=0.667$.
We have further confirmed this behavior by analyzing the cases $\alpha=1.4, 1.2, 0.8$ and 0.6.

\begin{figure}
\includegraphics[width=\columnwidth]{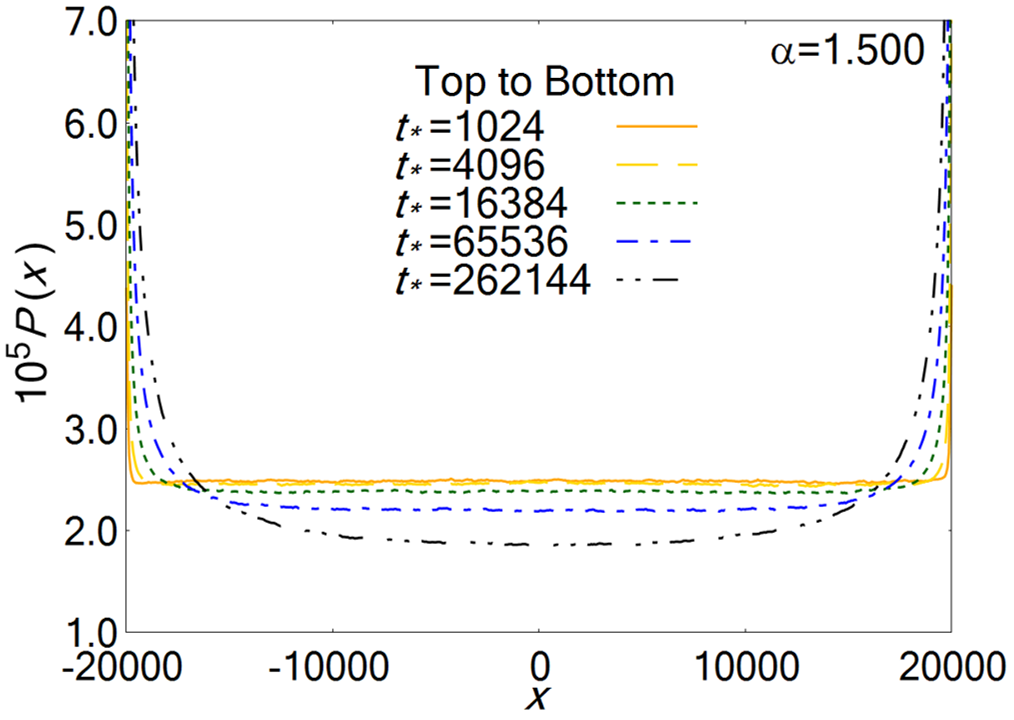}
\caption{Stationary probability density $P$ vs.\ position $x$ of exponentially tempered FBM for
interval length $L=40,000$, anomalous diffusion exponent $\alpha=1.5$, and several values of the
tempering time $t_*$. The data are averages over $2^{17} \approx 131,000$ time steps after the steady
state has been reached for $1.2 \times 10^6$ particles. The resulting relative statistical error
of $P$ is about $10^{-3}$.}
\label{fig:pdf_alpha150}
\end{figure}

\begin{figure}
\includegraphics[width=\columnwidth]{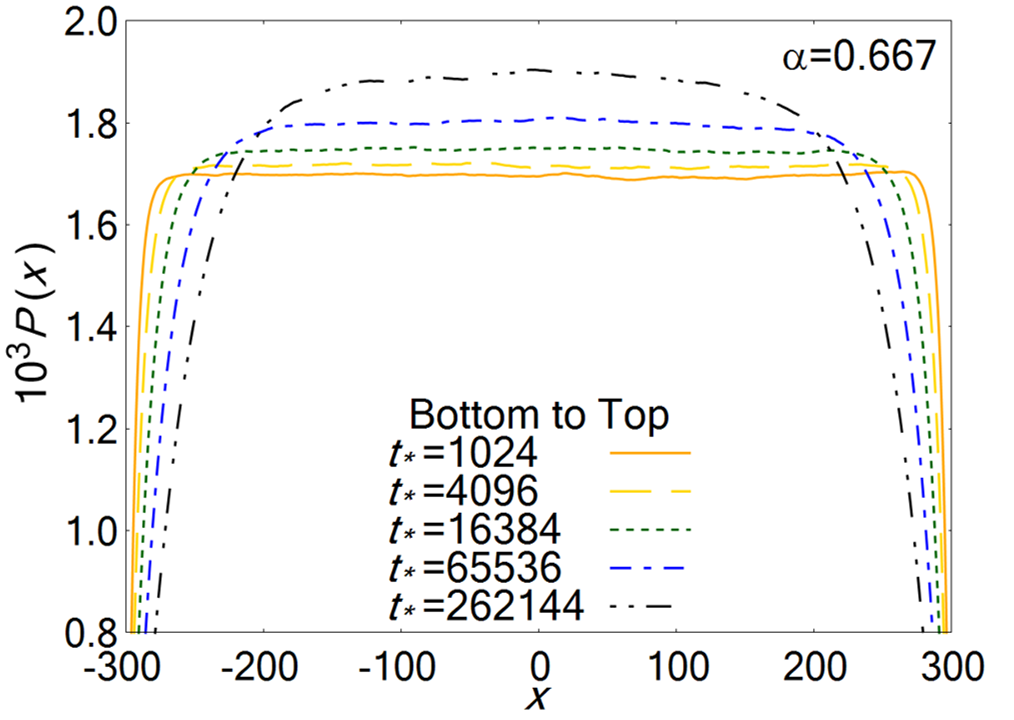}
\caption{Stationary probability density $P$ vs.\ position $x$ of exponentially tempered FBM for
interval length $L=600$, anomalous diffusion exponent $\alpha=0.667$, and several values of the
tempering time $t_*$. The data are averages over $2^{17} \approx 131,000$ time steps after the steady
state has been reached for $1.2 \times 10^6$ particles.}
\label{fig:pdf_alpha0667}
\end{figure}

In the following, we focus on the steady state reached at sufficiently long times and investigate its
probability density function. Figure \ref{fig:pdf_alpha150} presents an overview over the
stationary probability density $P(x)$ for the (superdiffusive) anomalous diffusion exponent $\alpha=1.5$
and several tempering times $t_*$.
The data show that particles accumulate close to the wall for all tempering times. The width
of the accumulation region decreases with decreasing tempering time because the long-time
correlations responsible for the accumulation are cut off at a distance $d_*$ from the wall,
defined by $d_*^2= 2 K t_*^\alpha$. For positions $x$ outside of the accumulation region of width $d_*$,
the stationary probability density is constant in agreement with the normal diffusion behavior at
times beyond $t_*$.
Analogous behavior is observed in the subdiffusive case $\alpha=0.667$,
as illustrated in Fig.\ \ref{fig:pdf_alpha0667}.
Here, particles are depleted close to the wall for all tempering times, and the width of the
depletion region varies with $t_*$ as above.

\begin{figure}
\includegraphics[width=\columnwidth]{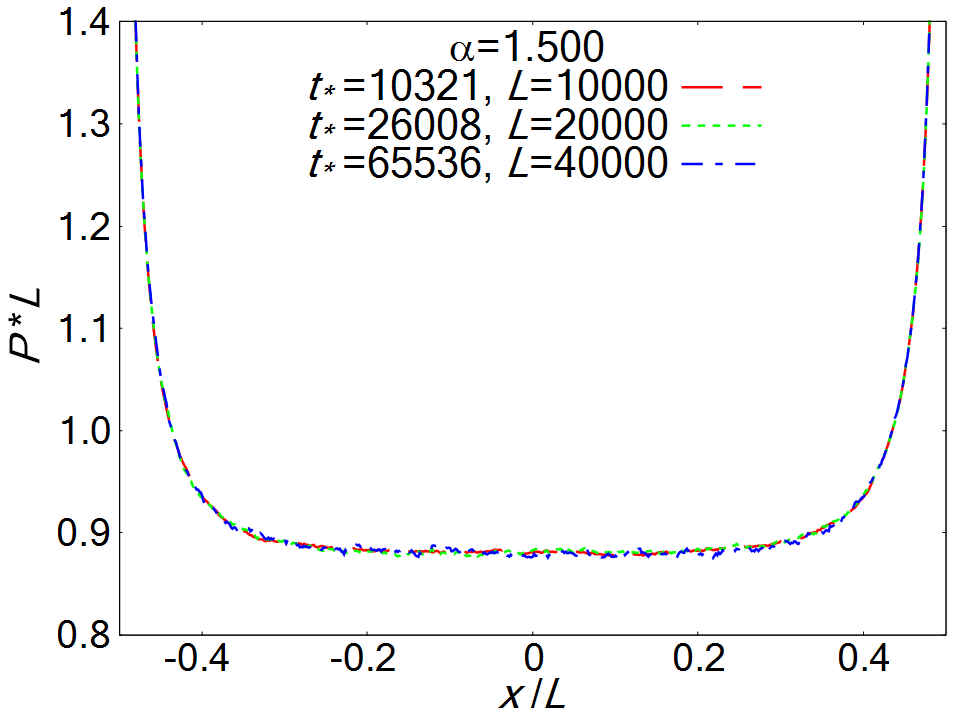}
\caption{Scaling plot of the stationary probability density of exponentially tempered FBM
showing $P L$ vs.\ position $x/L$ for $\alpha=1.5$. The interval lengths $L$ and tempering times $t_*$
have been chosen such that the second argument of the scaling function $Z_\alpha$ in eq.\
(\ref{eq:P_scaling_L_t*}) stays constant. The data are averages over $2^{17}$ time steps for
$1.2 \times 10^6$ particles.}
\label{fig:tempered scaling}
\end{figure}

The emergence of the cutoff distance $d_*$ as a new length scale suggests a generalization of
the scaling form (\ref{eq:P_scaling_L}) for untempered FBM to the tempered case. The stationary
probability density of exponentially tempered FBM with tempering time $t_*$ on an interval
of length $L$ is expected to fulfill the scaling form
\begin{equation}
P(x,L,t_*) = \frac 1 L Z_\alpha(x/L,t_*^{\alpha/2}/L)~.
\label{eq:P_scaling_L_t*}
\end{equation}
To verify that $P$ fulfills this scaling form, we have performed simulations at fixed $\alpha$ for several
interval lengths and adjusted the tempering times such that the second argument of the scaling
function $Z_\alpha$ stays constant. An example of this analysis is shown in Fig.\ \ref{fig:tempered scaling}
which confirms that the resulting stationary probability densities indeed collapse onto a single
master curve when plotted as $P L$ vs.\ $x/L$.

\begin{figure}
\includegraphics[width=\columnwidth]{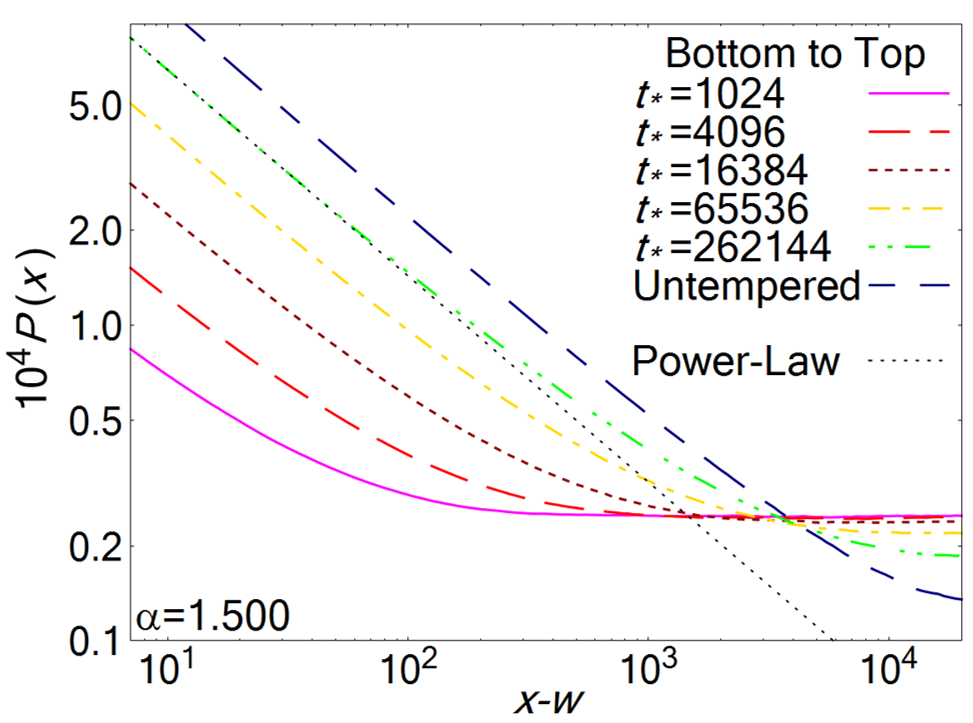}
\caption{Log-log plot of the stationary probability density $P$ vs.\ the distance $x-w$ from the left reflecting wall
($w=-L/2$) for exponentially tempered FBM, interval length $L=40,000$, anomalous diffusion exponent $\alpha=1.5$,
 and several values of the
tempering time $t_*$. The data are averages over $2^{17}$ time steps for $1.2 \times 10^6$ particles.
The dotted line is a power-law fit $P \sim (x-w)^\kappa$ using the same exponent
$\kappa=2/\alpha -2 = -2/3$ as applies to untempered FBM.}
\label{fig:log_alpha150}
\end{figure}

Let us now turn to the functional form of the stationary probability density close to the reflecting wall.
As the probability density of untempered FBM develops a power-law singularity $P \sim (x-w)^\kappa$
with $\kappa=2/\alpha -2 $ as function of the distance $x-w$ from the wall, we present in Fig.\ \ref{fig:log_alpha150} a
double-logarithmic plot (where power-laws are represented by straight lines) of $P$ near the left interval boundary
($w=-L/2$) vs.\ distance $x-w$ from the wall for $\alpha=1.5$.
The probability densities for all tempering times $t_*$ display power-law behavior sufficiently close to
the wall (for positions within their respective accumulation regions). The asymptotic behavior near the wall
can be fitted well by the same power law,  $P \sim (x-w)^\kappa$ with $\kappa=2/\alpha -2 = -2/3$, as holds for
untempered FBM. (As all curves become parallel for small $x-w$, this power law holds for all $t_*$.)
The behavior near the right interval boundary is completely analogous.

\begin{figure}
\includegraphics[width=\columnwidth]{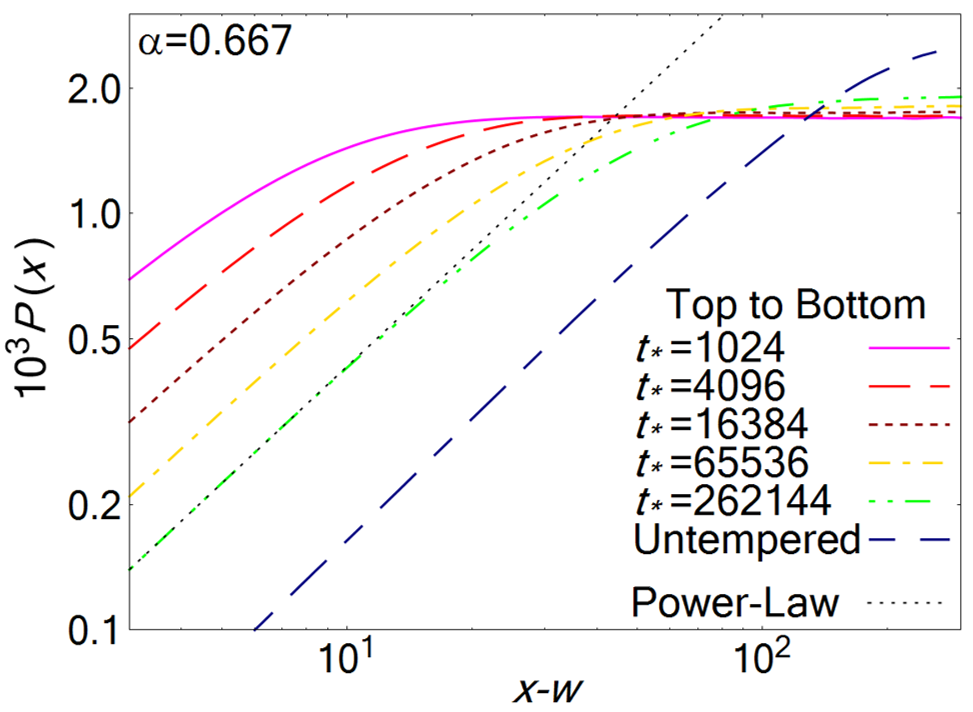}
\caption{Log-log plot of the stationary probability density $P$ vs.\ the distance $x-w$ from the left reflecting wall
($w=-L/2$) for exponentially tempered FBM, interval length $L=600$, anomalous diffusion exponent $\alpha=0.667$,
 and several values of the
tempering time $t_*$. The data are averages over $2^{17}$ time steps for $1.2 \times 10^6$ particles.
The dotted line is a power-law fit $P \sim (x-w)^\kappa$ using the same exponent
$\kappa=2/\alpha -2 = 1$ as applies to untempered FBM.}
\label{fig:log_alpha0667}
\end{figure}

Figure \ref{fig:log_alpha0667} presents the same analysis for the subdiffusive case of $\alpha=0.667$.
It demonstrates that the stationary probability density behaves as a power-law,
$P \sim (x-w)^\kappa$ with $\kappa=2/\alpha -2 = 1$, in the depletion region close to the wall
for all tempering times.

%%%%%%%%%%%%%%%%%%%%%%%%%%%%%%%%%%%%%%%%%%%%%%%%%%%%%%%%%%%%%%%%%%%%%%%%%%%%%%%%%%%%%%%%%%%%%%%%%%%%%%
\section{Results: power-law tempered fractional Brownian motion}
\label{sec:results_pow}
%%%%%%%%%%%%%%%%%%%%%%%%%%%%%%%%%%%%%%%%%%%%%%%%%%%%%%%%%%%%%%%%%%%%%%%%%%%%%%%%%%%%%%%%%%%%%%%%%%%%%%

As explained in Sec.\ \ref{subsec:tempering}, the properties of power-law tempered FBM, characterized by the
noise covariance (\ref{eq:FGN_cov_pow}), are more complex than
those of exponentially tempered FBM. Moreover, superdiffusive and subdiffusive FBM are affected by the tempering
in qualitatively different fashions.

\begin{figure}
\includegraphics[width=\columnwidth]{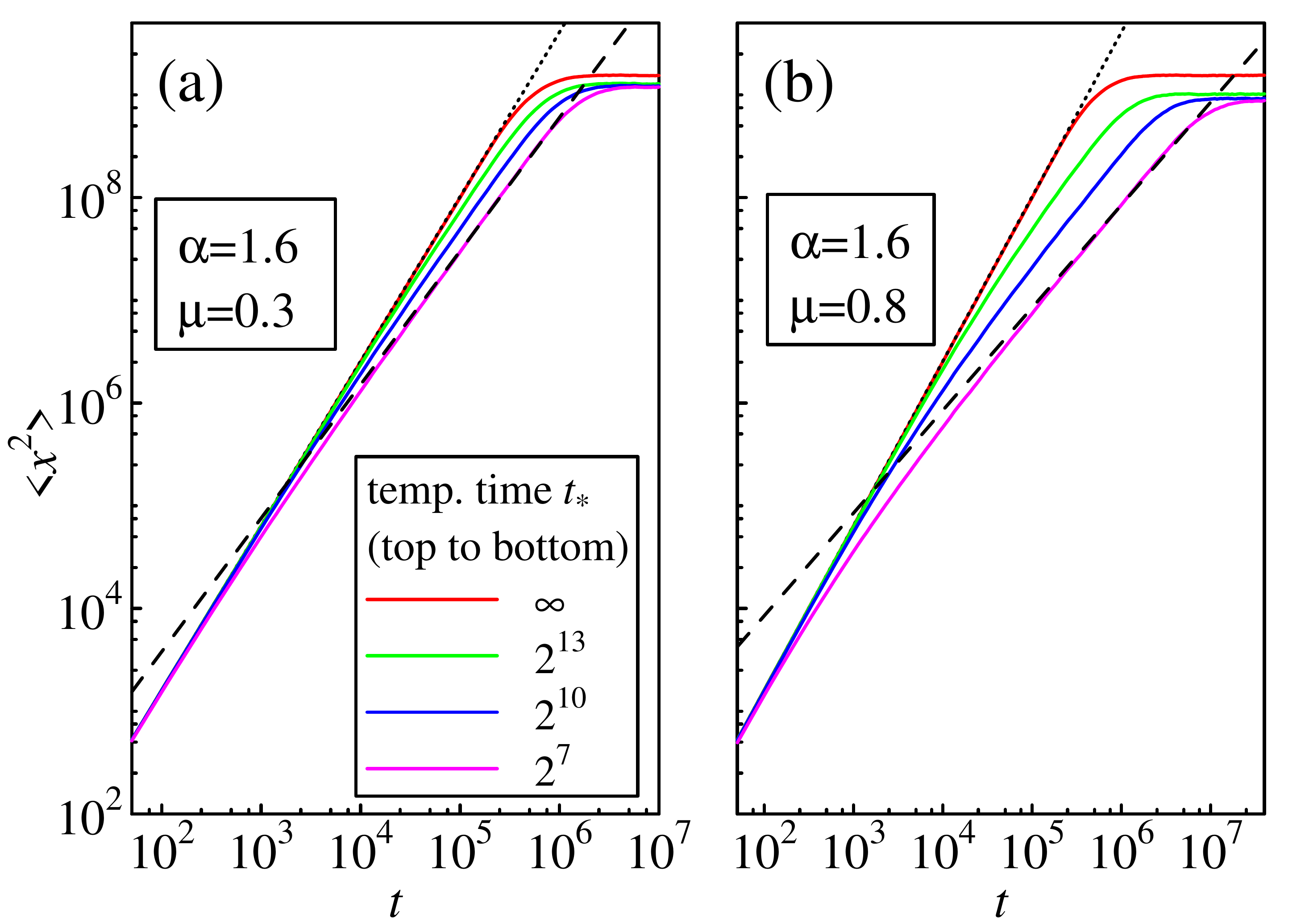}
\caption{Mean-square displacement $\langle x^2 \rangle$ vs.\ time $t$ of power-law tempered FBM for
interval length $L=10^5$ and several values of the tempering time $t_*$.
(a) Weak power-law tempering, $\alpha=1.6$, $\mu=0.3$.
(b) Strong power-law tempering, $\alpha=1.6$, $\mu=0.8$
The data are averages over 20,000  particles. The resulting relative statistical
error of $\langle x^2 \rangle$ is about $10^{-2}$, well below the line width.
The dotted lines are fits of the
early-time behavior to $\langle x^2 \rangle \sim t^\alpha$.
The dashed line is a fit to $\langle x^2 \rangle \sim t^{\alpha-\mu}$ in panel (a)
while it represents a fit to normal diffusion $\langle x^2 \rangle \sim t$ in panel (b).}
\label{fig:av_all_mu03_mu08}
\end{figure}

Let us start with the superdiffusive case ($1 < \alpha < 2$). If the tempering exponent $\mu$
fulfills the inequality $\alpha -\mu > 1$ (weak power-law tempering), the motion crosses over from anomalous
diffusion governed by $\langle x^2 \rangle \sim t^\alpha$ at times $t \ll t_*$ to anomalous diffusion
$\langle x^2 \rangle \sim t^{\alpha-\mu}$ for times $t \gg t_*$ \cite{MolinaGarciaetal18}.
The behavior of the mean-square displacement for weakly power-law tempered FBM on a finite
interval is illustrated in Fig.\
\ref{fig:av_all_mu03_mu08}(a) for $\alpha=1.6$ and $\mu=0.3$.
The data demonstrate two distinct anomalous diffusion regimes with exponents $\alpha$
and $\alpha -\mu$ before the mean-square displacement saturates when the particles have
spread over the interval.
For strong power-law tempering ($\alpha -\mu < 1$), in contrast, the motion for times
$t \gg t_*$ is normal diffusion. This can be seen in  Fig.\
\ref{fig:av_all_mu03_mu08}(b) which presents the mean-square displacement for
$\alpha=1.6$ and $\mu=0.8$. Note that the crossover from anomalous to normal
diffusion is much slower than in the case of exponential tempering, see Fig.\ \ref{fig:msd_alpha150}.

\begin{figure}
\includegraphics[width=\columnwidth]{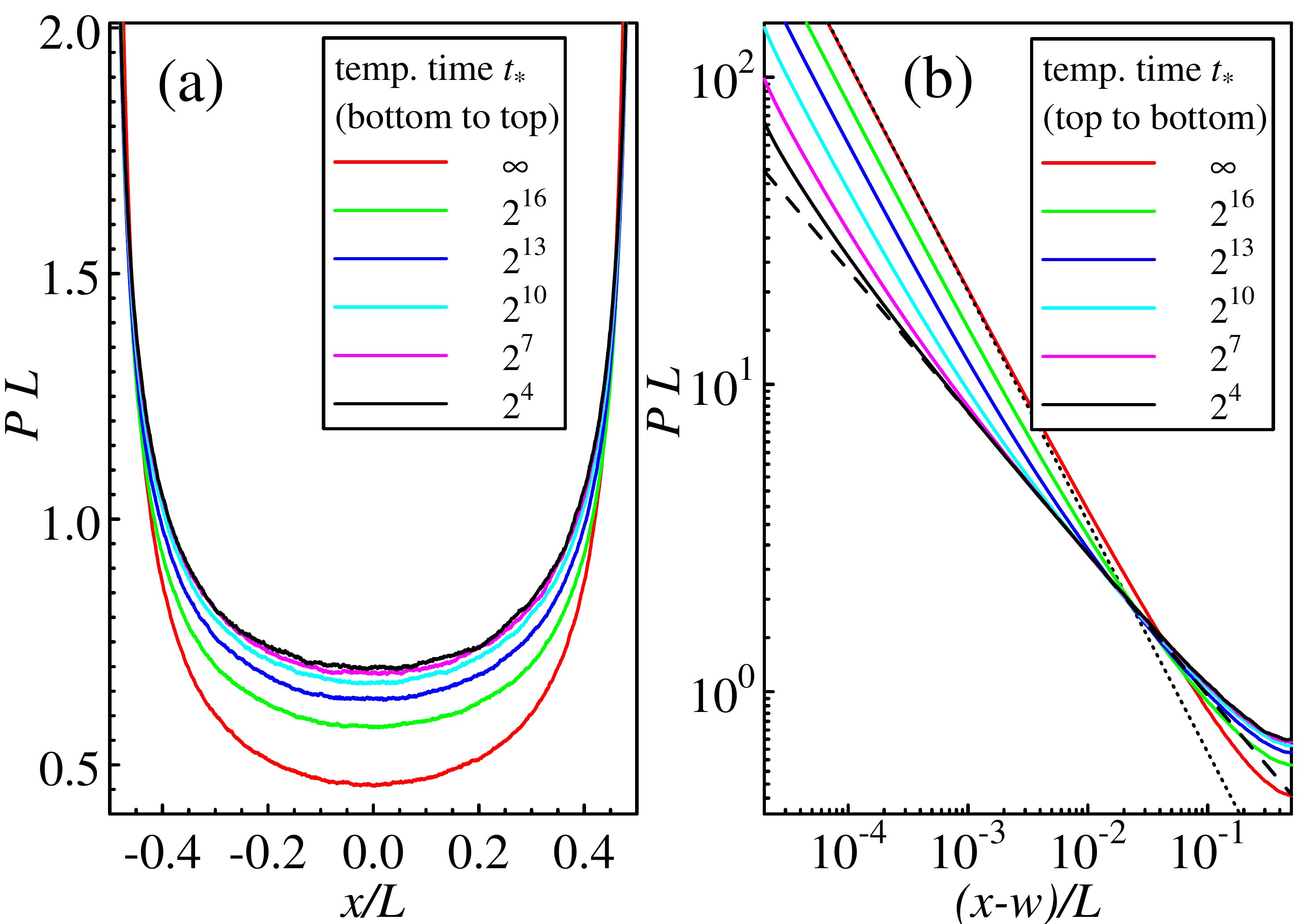}
\caption{(a) Scaled stationary probability density $P L$ vs.\ scaled position $x/L$ of power-law tempered FBM for
interval length $L=10^5$, anomalous diffusion exponent $\alpha=1.6$, tempering exponent $\mu=0.3$, and several values
of the tempering time $t_*$. The data are averages over $2^{25} \approx 33$ million time steps after the steady
state has been reached for 20,000 particles. (b) Log-log plot of the scaled stationary probability density $P L$
vs.\ the scaled distance from the wall $(x-w)/L$. The dotted line is a power-law fit $P \sim (x-w)^\kappa$ of the
untempered data using $\kappa=2/\alpha -2 = -0.75$. The dashed line is a power-law fit of the preasymptotic
behavior for $t_*=2^4$ using the exponent $\kappa=2/(\alpha -\mu) -2 \approx -0.462$.}
\label{fig:pdf_log_alpha16_mu03}
\end{figure}

We now discuss the stationary probability density for superdiffusive power-law tempered FBM on a finite
interval.
Figure \ref{fig:pdf_log_alpha16_mu03}(a) presents an overview of the stationary probability density
for $\alpha=1.6$ and $\mu=0.3$, i.e., for a weak tempering situation.
In contrast to the behavior of exponentially tempered FBM (see Fig.\ \ref{fig:pdf_alpha150}), the probability
density does not become flat away from the reflecting walls, even for the shortest tempering time of only $t_*=2^4$.
This reflects the fact that the motion does not cross over to normal diffusion but remains superdiffusive
beyond $t_*$. Figure \ref{fig:pdf_log_alpha16_mu03}(b) analyzes the functional form of the probability density $P$
near the reflecting wall. The data demonstrate that $P$ follows the power law $P \sim (x-w)^\kappa$ with
$\kappa=2/\alpha -2$ asymptotically close to the wall. Outside the asymptotic region of width $d_*= (2 K t_*^\alpha)^{1/2}$,
the behavior is governed by the anomalous diffusion exponent $\alpha - \mu$.
Assuming that the condition $d_* \ll L$ is fulfilled, we therefore expect a well-defined preasymptotic region
$d_* \ll x \ll L$ in which the probability density follows a power law $P \sim (x-w)^\kappa$, but with exponent
$\kappa = 2/(\alpha-\mu)-2$. This behavior is indeed observed in Fig.\ \ref{fig:pdf_log_alpha16_mu03}(b).

We have performed an analogous analysis for the strongly power-law tempered case of $\alpha=1.6$ and $\mu=0.8$.
In agreement with the fact that the motion crosses over to normal diffusion for times beyond $t_*$, the properties
of the stationary probability density qualitatively resemble those of exponentially tempered FBM
(Figs.\  \ref{fig:pdf_alpha150} and \ref{fig:log_alpha150}) rather than those of weakly power-law tempered FBM.
Specifically, $P$ follows the power law $P \sim (x-w)^\kappa$ with
$\kappa = 2/\alpha-2$ asymptotically close to the wall, but outside of the asymptotic
region of width $d_*$, the probability density approaches the constant behavior
expected for normal diffusion. As in the case of the mean-square displacement [Fig.\ \ref{fig:av_all_mu03_mu08}(b)],
the crossover between the anomalous and normal diffusion regimes is much slower than in the exponentially tempered case.

So far, our discussion of power-law tempered FBM has focused on the superdiffusive case. We now turn to
subdiffusive power-law tempered FBM. The discussion in Sec.\ \ref{subsec:tempering} emphasized that the
subdiffusive behavior of FBM with $\alpha <1$ is the result of the perfect anticorrelations of the
corresponding fractional Gaussian noise, encoded in the relation $\sum_n C_n = 0$ for the noise covariance.
This is equivalent to a vanishing of the covariance Fourier component $\tilde C(0)$.
These anticorrelations are fragile, however, as any modification of the noise covariance function generically
leads to a violation of the relation $\sum_n C_n = 0$ unless the covariance is fine tuned.
More specifically, the power-law tempered noise with covariance (\ref{eq:FGN_cov_pow}) violates the perfect
anticorrelation condition for all $\mu$ and $t_*$. Consequently, power-law tempered subdiffusive FBM
is expected to cross over from anomalous diffusion for times below $t_*$ to normal diffusion at longer
times.\footnote{Naively, one might have expected a crossover between two anomalous diffusion regimes,
characterized by anomalous diffusion exponent values $\alpha$ (for times below $t_*$) and $\alpha-\mu$
(for times above $t_*$). Because of the fragility of the anticorrelations in subdiffusive FBM, this is
not the case.}

\begin{figure}
\includegraphics[width=\columnwidth]{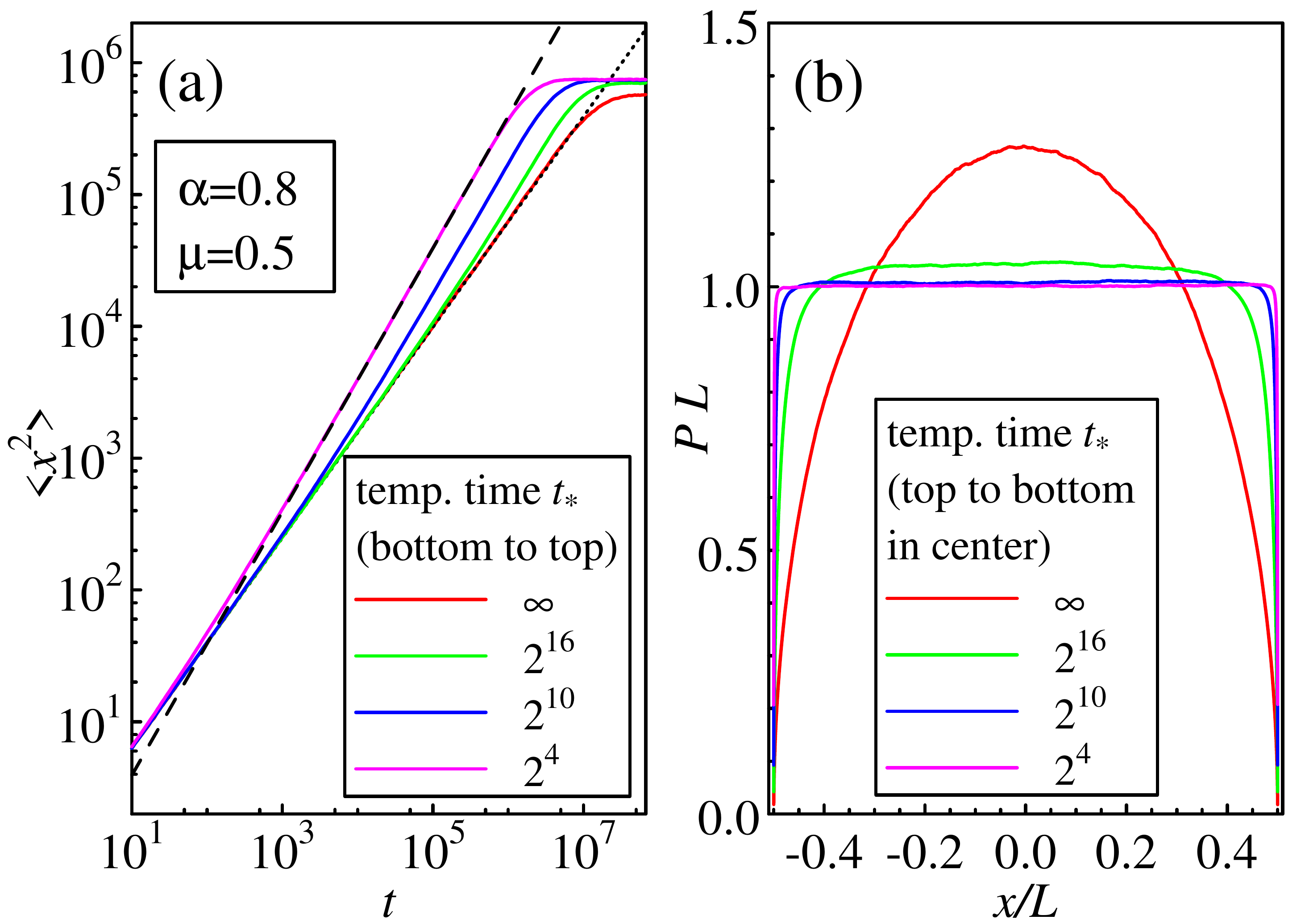}
\caption{(a) Mean-square displacement $\langle x^2 \rangle$ vs.\ time $t$ of power-law tempered FBM for
$\alpha=0.8$, $\mu=0.5$, interval length $L=3000$ and several values of the tempering time $t_*$.
The data are averages over 50,000  particles. The dotted line is a fit of the
early-time behavior to $\langle x^2 \rangle \sim t^\alpha$.
The dashed line is a fit of the behavior after $t_*$ to normal diffusion $\langle x^2 \rangle \sim t$.
(b) Scaled stationary probability density $P L$ vs.\ scaled position $x/L$ for the same parameters as in
panel (a). The data are averages over $2^{25} \approx 33$ million time steps after the steady
state has been reached.}
\label{fig:av_log_all_gamma12_mu05}
\end{figure}

Figure \ref{fig:av_log_all_gamma12_mu05}(a) presents the time evolution of the mean-square displacement for
$\alpha=0.8$, $\mu=0.5$ and several $t_*$; the data confirm this expectation.
The (scaled) stationary probability density  for the same stochastic processes is presented in Fig.\
\ref{fig:av_log_all_gamma12_mu05}(b)  which shows that the probability density $P$ goes to zero
at the reflecting wall, as in the untempered case. Power-law fits demonstrate that $P$ follows the
same asymptotic behavior, $P \sim (x-w)^\kappa$ with $\kappa=2/\alpha-2$ as in the untempered case.
Outside the asymptotic
region of width $d_*= (2 K t_*^\alpha)^{1/2}$, $P$ approaches the constant behavior
expected for normal diffusion.

%%%%%%%%%%%%%%%%%%%%%%%%%%%%%%%%%%%%%%%%%%%%%%%%%%%%%%%%%%%%%%%%%%%%%%%%%%%%%%%%%%%%%%%%%%%%%%%%%%%%%%
\section{Conclusions}
\label{sec:conclusions}
%%%%%%%%%%%%%%%%%%%%%%%%%%%%%%%%%%%%%%%%%%%%%%%%%%%%%%%%%%%%%%%%%%%%%%%%%%%%%%%%%%%%%%%%%%%%%%%%%%%%%%

In summary, we have employed large-scale computer simulations to study tempered FBM \cite{MolinaGarciaetal18}, a
stochastic process with long-time power-law correlations that are cut off at some mesoscopic time scale,
the tempering time $t_*$. Specifically, we have analyzed the behavior of tempered FBM confined
to a finite one-dimensional interval by means of reflecting walls in order to understand how the
tempering of the correlations affects the unusual accumulation and depletion effects recently
observed for (untempered) reflected FBM.

The motion of particles that start at the center of the interval features three distinct time
regimes (assuming the interval length is sufficiently large and/or the tempering time is sufficiently small).
At times below $t_*$, the particles spread exactly as they would for untempered FBM. Beyond $t_*$, the
particles continue to spread but the motion changes qualitatively due to the cutoff of the correlations.
At the longest times, when the particles have spread over the entire interval, the particle distribution
reaches a stationary state.

The character of the stochastic process beyond $t_*$ depends on the type of the tempering. For a hard
exponential cutoff of the correlations, the motion crosses over to normal diffusion. For the softer
power-law tempering, the behavior is more complex and depends on the values of $\alpha$ and $\mu$.
The motion beyond $t_*$ is of normal diffusion type if the underlying FBM is either superdiffusive with
$\alpha -\mu <1$ or subdiffusive (for any subdiffusive $\alpha$ and $\mu>0$). For superdiffusive
power-law tempered FBM with $\alpha-\mu > 1$, in contrast, the motion beyond $t_*$ is anomalous diffusion
with a reduced anomalous diffusion exponent value of $\alpha -\mu$.

The main focus of the present paper has been on the stationary probability density that the stochastic process reaches
after sufficiently long times. Our simulation results demonstrate that tempered FBM features the same
accumulation and depletion effects close to a reflecting wall as untempered FBM. More specifically, particles
accumulate near the wall in the superdiffusive case but are depleted at the wall in the subdiffusive case.
Asymptotically close to the wall, the functional form of the stationary probability density of tempered
FBM is governed by the same power-law singularity $P \sim (x-w)^\kappa$ with $\kappa = 2/\alpha-2$ as untempered
FBM ($x-w$ represents the distance from the wall). However, due to the cutoff of the correlations,
this power-law behavior is restricted to a region of finite width $d_*= (2 K t_*^\alpha)^{1/2}$ near the wall.
Outside of this region, the probability density becomes flat in the cases where the motion beyond $t_*$ is of
normal diffusion type. The most interesting case occurs for superdiffusive power-law tempered FBM
with $\alpha -\mu >1$. Here, the probability density features two power-law regimes with $\kappa$
values $\kappa = 2/\alpha-2$ (asymptotically close to the wall) and $\kappa = 2/(\alpha-\mu)-2$
(for $d_* \ll |x-w| \ll L$).

We also found that the tempering of the correlations introduces the new length scale $d_*$ and thus leads to the generalized
scaling form (\ref{eq:P_scaling_L_t*}) of the stationary probability density. Our numerical data fulfill
this scaling form with high accuracy.

Let us now put our results into a broader perspective. In the present work we have considered tempered
FBM confined to a finite interval by two reflecting walls. Instead, one could also consider a situation with
only a single reflecting wall and introduce a bias (nonzero mean of the increments) towards the wall
as was done for untempered FBM in Ref.\ \cite{WadaWarhoverVojta19}. We expect that the behavior of such a system
close to the wall is qualitatively identical to the behavior found in the present paper.

It is also interesting to consider a generalized Lange\-vin equation driven by the same tempered fractional Gaussian
noise as the tempered FBM studied in the present paper \cite{MolinaGarciaetal18}. A key question is whether the
probability density
of such a Langevin equation confined to a finite interval also shows accumulation and or depletion effects close
to the confining walls. If the generalized Langevin equation fulfills the fluctuation-dissipation theorem
(which connects the noise covariance and the damping kernel), the stationary state is expected to be a thermal
equilibrium state which has a flat probability density independent of the values of $\alpha$ and $\mu$.
(For the untempered fractional Langevin equation, this absence of accumulation and depletion effects was recently
observed in simulations \cite{VojtaSkinnerMetzler19}.) This highlights that the nonequilibrium nature of FBM
(tempered or untempered) is responsible for the accumulation and depletion effects near a reflecting wall.

We emphasize that the notion of tempering the fractional Gaussian noise, as introduced in Ref.\ \cite{MolinaGarciaetal18}
and employed in the present paper, differs fundamentally from a model proposed by Meerschaert and Sabzikar
\cite{MeerschaertSabzikar13} in which exponential tempering factors are introduced directly into Mandelbrot's definition
\cite{MandelbrotVanNess68} of FBM. That process does not describe the crossover from anomalous diffusion to
normal diffusion. Instead, its mean-square displacement approaches a constant in the long-time limit,
i.e., it describes a confined motion \cite{ChenWangDeng17,MolinaGarciaetal18}. A Langevin equation driven by the
corresponding noise leads to ballistic long-time behavior, very different from the processes considered in the present
paper.

Finally, we point out that the tempering of the correlations provides a powerful tool in applications in which a
stochastic process is used to model experimental data. For example, FBM was recently put forward as a
model to explain the spatial distribution of serotonergic fibers in vertebrate brains
\cite{JanusonisDetering19,JanusonisDeteringMetzlerVojta20}. Despite  the limited ``neurobiological input'',
the model captures important aspects of the highly nonuniform distributions of these fibers throughout
the brain. Tempering will permit further refinements of the model to better represent the observed
fiber densities. We expect similar advantages in many other applications.

%%%%%%%%%%%%%%%%%%%%%%%%%%%%%%%%%%%%%%%%%%%%%%%%%%%%%%%%%%%%%%%%%%%%%%%%%%%%%%%%%%%%%%%%%%%%%%%%%%%%%%%%%%%%%%%%
\begin{acknowledgements}
This work was supported in part by a Cottrell SEED award from Research Corporation and by
the National Science Foundation under Grant Nos.\ DMR-1828489 and OAC-1919789.
The simulations were performed on the Pegasus and Foundry clusters at Missouri S\&T.
We acknowledge helpful discussions with Ralf Metzler and Skirmantas Janusonis.
\end{acknowledgements}

{\small \noindent
{\bf Author Contribution Statement}
T.V. conceived and coordinated the study. Z.M. and S.H. performed the computer simulations and analyzed the data.
Z.M. and T.V. created the figures. T.V. wrote the manuscript.\par
}
\def\bibfont{\footnotesize}

%\bibliographystyle{spphys-tv}
\bibliographystyle{apsrev4-1}
\bibliography{../00bibtex/rareregions}   % name your BibTeX data base

\end{document}